# Cooperative Spin Caloritronic Devices

Fengjun Zhuo[1,2], Z. Z. Sun[1,*] and Jian-Hua Jiang[1,#]

[1] *College of Physics, Optoelectronics and Energy & Jiangsu Key Laboratory of Thin Films, Soochow University, Suzhou, Jiangsu 215006, China*

[2] *Jiangsu Key Laboratory for Carbon-Based Functional Materials & Devices, Institute of Functional Nano & Soft Materials (FUNSOM), Soochow University, Suzhou, Jiangsu 215123, China*

We report a concept of thermoelectric devices, cooperative spin caloritronics device (CSCD), where cooperation between two or more energy channels such as spin, charge and heat currents can significantly enhance energy efficiency of spin caloritronic devices. We derive the figure of merit and the maximum efficiency due to cooperative effect in analytic forms for a CSCD. Cooperative effects significantly improve the figure of merit and the maximum efficiency in spin caloritronic systems with multiply couplings effects. Several examples of CSCDs, including electrical and thermal current induced DW motion, spin-thermoelectric power generator and spin-thermoelectric cooling/heating, are studied to illustrate the usefulness of the cooperative effect. We compare the efficiency of CSCD with several recently proposed spin caloritronic devices. Our scheme provides a novel route to seek high performance materials and structures for future spin caloritronic devices.

PACS numbers: 72.20.Pa, 72.20.My, 75.78.Jp, 85.80.-b

*Introduction.* In the past few years, manipulation of magnetization and magnetic textures such as domain walls (DW) and skyrmions in ferromagnetic (FM) nanostructures has attracted a lot of attention because of fundamental interest and potential impacts on data storage devices and logic operations [1-5]. Interplay between electronic spin, charge, and magnetization offers a promising physical mechanism for such manipulation. Especially, current-induced DW motion [6-11] or skyrmions [12, 13] along highly conducting magnetic nanowires promises the development of novel spintronic devices with high density, performance and endurance at a very low cost per bit, such as racetrack memory [4]. However, extensive experimental [4, 14] and theoretical [6-8] studies have shown that the critical current density to drive the motion of the conventional magnetic DW in FM nanostructures is on the order of $10^5 - 10^8$ A cm$^{-2}$. Joule heating in such a high density information processing scheme becomes a serious issue because of the large current density which is necessary to overcome the pinning.

Recently it was proposed theoretically and verified experimentally that heat current can also serve as an efficient way to drive the motion of DW [15-22] and skyrmions [23-25]. It may be possible to exploit waste Joule heat to assist current-driven magnetic patterns motion for information processing. Alongside with electrical current and spin current, heat can be conducted to designated regions to achieve DW manipulation effectively. Such realizations lead to prosperous researches on spin caloritronics [26-29], an emerging field to study the interaction between spin, charge and heat currents, and magnetization in magnetic materials and structures. Pioneering researches have uncovered abundant physical mechanisms, such as electron-magnon, phonon-magnon, and charge-spin couplings that explain the versatile phenomena in spin caloritronics systems [15-21, 30, 31]. Those couplings provide new ways to manipulate magnetic textures for information storage and processing. However, as for the situation of current-induced DW motion, energy efficiency in those couplings are still very low, which is one of the main challenges for spin-caloritronic applications [32-36].

In this Rapid Communication, we propose a novel concept of cooperative spin caloritronics device (CSCD) where cooperation between two or more energy channels can significantly enhance energy efficiency of spin caloritronic devices. Theoretical foundation of such cooperative effects is established in Ref. [37] based on Onsager's theory of irreversible thermodynamics. A typical CSCD can be DW motion driven by coexisting electrical and heat currents. We show that cooperation between electrical and thermal currents induced DW motion can greatly improve the energy efficiency, surpassing the maximum achievable efficiency for DW motion induced solely by electrical or thermal current. Other CSCDs include spin-thermoelectric power generator and spin-thermoelectric cooling/heating. Our scheme provides a new route to significantly enhance energy efficiency and hence considerably reduce Joule heating for future advanced magnetic information storage and information processing.

*Basic theoretical framework.* Onsager's theory of irreversible thermodynamics establishes a general form to study nonequilibrium phenomena in thermodynamic systems [38]. Like in classical systems with balanced friction and driving forces and moving in constant velocity, thermodynamic systems under external forces derives motions ("currents") at steady states. The relation between the forces $\vec{\mathcal{F}}$ and currents $\vec{\mathcal{J}}$ is generally written as [38-40], $\vec{\mathcal{J}} = \widehat{M}\vec{\mathcal{F}}$ or $\mathcal{J}_n = \sum_k M_{nk}\mathcal{F}_k$, where the index $n$ ($k$) numerates all currents (forces), and $\widehat{M}$ is called the Onsager matrix. When the forces are not too strong, the dependence of $\widehat{M}$ on the forces can be ignored. Cross-correlated responses (e.g., thermoelectric effect) allow conversion from the input energy to the output energy (e.g., thermal to electrical energy conversion). In general, a thermodynamic machine realizes its function via consuming the input energy and converting this energy into the output work/energy to achieve certain functionalities. According to the theory of irreversible thermodynamics [37-38], there are an equal number of forces and currents.

An important aspect of the performance of a machine is its energy efficiency. High energy-efficiency machine is demanded for future society not only to reduce energy cost, but also because damage of materials can be reduced if heating due to irreversible dissipation is reduced. It is hence crucial to improve the energy efficiency of functional materials and machines made of these materials. In practical applications, the first target is to find out the optimal energy efficiency and the condition that realizes the optimal energy efficiency for the functional materials/systems [41-43]. A general theory [37] was developed to fulfill this target for thermodynamic systems with arbitrary Onsager matrix (that may describe complex responses to multiple forces). We repeat the derivation of the optimal efficiency in the supplementary materials [44] where we also give other forms of the results beside that was presented in Ref. [37]. For symmetric Onsager matrices (i.e., time-reversal symmetric systems), $\widehat{M}_{II} = \widehat{M}_{II}^T$, $\widehat{M}_{OI} = \widehat{M}_{IO}^T$, and $\widehat{M}_{OO} = \widehat{M}_{OO}^T$. The maximum energy efficiency of the systems with a symmetric Onsager matrix is

$$\phi_{max} = \frac{\sqrt{\xi+1}-1}{\sqrt{\xi+1}+1} \qquad (1)$$

where $\xi = \frac{\lambda}{1-\lambda}$ is the figure of merit and $\lambda$ is called the "degree of coupling" [41, 47]. Here $\lambda$ is the largest eigenvalue of $\widehat{\Lambda} = \widehat{M}_{II}^{-1/2} \widehat{M}_{IO} \widehat{M}_{OO}^{-1} \widehat{M}_{OI} \widehat{M}_{II}^{-1/2}$ (termed as the "coupling matrix") [44-49].

*Electrical and thermal current induced DW motion.* It has been shown that electrical and thermal current induced DW motion in a magnetic nanowire (Fig. 1(a)) can be described by a phenomenological linear-response equation $\vec{\mathcal{F}} = \widehat{M}\vec{\mathcal{J}}$ [16, 50], where

$$\vec{\mathcal{J}} = (\mathcal{J}_c, \mathcal{J}_Q, \mathcal{J}_w,)^T, \quad \vec{\mathcal{F}} = (\Delta V, \Delta T/T, 2AM_s H_{ext})^T, \qquad (2a)$$

$$\widehat{M} = \begin{pmatrix} M_{cc} & M_{cQ} & M_{cw} \\ M_{cQ} & M_{QQ} & M_{Qw} \\ M_{cw} & M_{Qw} & M_{ww} \end{pmatrix}. \qquad (2b)$$

The three thermodynamic currents are the electrical current $J_c$, the thermal current $J_Q$, and the velocity of DW motion $\mathcal{J}_w = \dot{r}_w$ where $r_w$ is the center of the DW. The three thermodynamic forces that induce the currents are the voltage $\Delta V = (\mu_h - \mu_c)/e$ with $\mu_h$ ($\mu_c$) being the electrochemical potential of the hot (cold) terminal, the temperature difference $\Delta T/T = (T_h - T_c)/T$ with $T_h$ ($T_c$) being the temperature of the hot (cold) terminal, and the external magnetic field $H_{ext}$. Following Ref. [50], the coefficients of the linear-response matrix can be written as follows: $M_{cc} = R$, $M_{QQ} = R/LT^2$, $M_{ww} = \frac{2\mu_0 \alpha A M_s}{\Delta \gamma}$, $M_{cQ} = SR/LT$, $M_{cw} = \frac{\hbar}{e\Delta}p\beta$ and $M_{wQ} = \frac{\hbar}{e}\frac{1}{\Delta LT}(S'\beta' - Sp\beta)$. Here $R = \frac{l}{\sigma A}$ is electrical resistance of the device where $\sigma$ is the electrical conductivity, $l$ and $A$ are the length and area of the device, respectively. For a nanowire system as illustrated in Fig. 1, we take $l = 1\mu m$ and $A = 100 nm^2$ for our calculation. $L = 2.443 \times 10^{-8}\, W\Omega K^{-2}$ is the Lorenz number for metals. $T = 300K$ is the room temperature. We chose the material parameters that are close to those of the permalloy, viz., the saturation magnetization $M_s = 860 \times 10^3 A/m$, the DW width $\Delta = 100 nm$, the Gilbert damping $\alpha = 0.01$, the electrical conductivity $\sigma = 10^5 (\Omega cm)^{-1}$, and the Seebeck coefficient $S = 100\mu V/K$. $\mu_0$ is the vacuum permeability, $e$ is the electron charge, and $\gamma$ is the gyromagnetic ratio. Microscopically, the spin polarization, the Seebeck coefficient and the spin Seebeck coefficient are given by $P = \langle s_z \rangle$, $S = \frac{\langle E \rangle}{eT}$ and $S' = \frac{\langle E s_z \rangle}{eT}$. We have set the energy zero to be the (equilibrium) chemical potential, i.e., $\mu \equiv 0$, $s_z = 1$ or $-1$ for spin up and down, respectively. The average here is defined as $\langle \mathcal{O} \rangle = \sigma^{-1} \int dE \left(-\frac{\partial n_F}{\partial E}\right) \sum_s \sigma^{(s)}(E) \mathcal{O}$, where $\sigma^{(s)}(E)$ ($s = \uparrow, \downarrow$) is the spin- and energy-dependent conductivity. $\sigma = \int dE \left(-\frac{\partial n_F}{\partial E}\right) \sum_s \sigma^{(s)}(E)$ is the electrical conductivity. $n_F = 1/\left[\exp\left(\frac{E}{K_B T}\right) + 1\right]$ is the Fermi distribution of the carrier. The relationships presented here are the generalized Mott relations for spin-caloritronic systems. The $\beta$ and $\beta'$ terms are regarded as crucial in understanding magnetic DW dynamics driven by electrical and thermal currents [16]. Although $\beta$ and $\beta'$ can generally be different, in the following estimation we will take $\beta = \beta' = 0.1$.

The maximum efficiency is determined by the figure of merit and the degree of coupling according to Eq. (1). As shown in the supplementary materials[44], a neat

way to express the figure of merit is to introduce the following dimensionless coefficients

$$q_{ij} = \frac{M_{ij}}{\sqrt{M_{ii}M_{jj}}}. \tag{3}$$

The above coefficient represents the degree of coupling [37,46,48] for energy conversion between the $i^{-th}$ channel and the $j^{-th}$ channel. For example, $q_{cQ}$ represents the degree of coupling between electrical and thermal energy [37,48]. The figure of merit for electrical current-induced DW motion is $\xi_{cw} = \frac{q_{cw}^2}{1-q_{cw}^2}$, and that for thermal current-induced DW motion is $\xi_{Qw} = \frac{q_{Qw}^2}{1-q_{Qw}^2}$. The second law of thermodynamics requires that $|q_{ij}| \leq 1$, so that the maximum efficiency is bounded from above to ensure $\phi_{max} \leq 100\%$. The figure of merit of the cooperative DW motion induced by the concurrent electrical and thermal currents is

$$\xi = \frac{1-q_{cQ}^2}{1-q_{cw}^2-q_{Qw}^2-q_{cQ}^2+2q_{cw}q_{Qw}q_{cQ}} - 1 , \tag{4}$$

which determines the maximum energy efficiency through Eq. (1). It can be proved that the cooperative figure of merit $\xi$ is always *larger* than (or, at least, equal to) $\xi_{cw}$ and $\xi_{Qw}$ (see Supplemental Material [44]). This is because the maximum efficiency given by the figure of merit in Eq. (4) is the global maximum of the efficiency, while $\xi_{cw}$ and $\xi_{Qw}$ only give the (conditional) maximum efficiency without heat or electrical current, respectively.

Fig. 1(b) demonstrates the energy efficiency as a function of the electrical and heat currents for a typical case. Specifically, the energy efficiency, $\phi$, as a function of the ratio of the input currents to the output current, $\mathcal{I}_c/\mathcal{I}_w$ and $\mathcal{I}_Q/\mathcal{I}_w$, is plotted. We set $p = 0.5$, $S = 100 \mu V/K$ and $S' = -80 \mu V/K$. These two currents can be of the same sign, or the opposite sign depending on the directions of the electrochemical potential gradient and the temperature gradient. Here we choose negative temperature gradient (along the $x$ direction) and vary the direction of the electrochemical potential gradient. The down-triangle (up-triangle) point represents the maximal energy efficiency for the magnetic DW motion driven solely by the thermal (electrical) current. The rhombus point represents the global maximum efficiency for the magnetic DW motion induced by the concurrent electrical and thermal currents. The cooperative effect is clearly manifested by the fact that the global maximum efficiency is much greater than the optimal efficiency's for the DW motion driven by only one of the currents, electrical or thermal current.

The enhancement of the maximum energy efficiency due to cooperative effects, measured by $\frac{\phi_{max}}{max(\phi_{cw},\phi_{Qw})}$, as a function of $P$ and $S'$ is plotted in Fig. 2(a). The energy efficiency is significantly improved by the cooperative effect when $S'/(100\mu V/K)$ is approximately negative twice of the spin polarization $P$. Fig. 2(b) shows the enhancement of the maximum energy efficiency as a function of the thermoelectric coupling coefficient $q_{cQ} = S/\sqrt{L}$ when the electrical and thermal current-induced DW motion coefficients, $q_{cw}$ and $q_{Qw}$, are set as constants (i.e., $S'\beta' - Sp\beta$ is fixed to be constant). Counterintuitively, although the thermoelectric coupling coefficient $q_{cQ}$ has nothing to do with the optimal energy efficiency of the electrical (or thermal) current induced DW motion, it has strong effects on the maximum efficiency for the magnetic DW motion driven by coexisting electrical and thermal currents. As already manifested in Eq. (4), the global maximum efficiency depends on the thermoelectric coupling coefficient $q_{cQ}$. Hence tuning $q_{cQ}$ can help improving the maximum efficiency. Fig. 2(b) shows that the dependence of the efficiency enhancement factor, $\frac{\phi_{max}}{max(\phi_{cw},\phi_{Qw})}$, on the thermoelectric coupling coefficient $q_{cQ}$ is not monotonic.

The non-monotonic behavior of the enhancement factor, $\frac{\phi_{max}}{max(\phi_{cw},\phi_{Qw})}$, can be understood via Eq. (4), since the optimal efficiency has a one-to-one correspondence to the figure of merit. We emphasize two important aspects of the cooperative effect. First, the magnetic DW motion induced by the electrical and the thermal currents can be of the same direction, leading to constructive interplay between the two driving factors. If their directions are opposite, however, there will be destructive interplay between the two. Second, entropy production that limits the maximum efficiency, has contribution from all processes, including electricity and heat to magnetic energy conversion, as well as the conversion between electricity and heat energy. Tuning the $q_{cQ}$ modifies the entropy production associated with the energy conversion between electricity and heat, and hence affects the maximum energy efficiency. However, the cooperative maximum efficiency is always *greater* than or equal to the maximum efficiency for magnetic DW motion driven by the electrical current or the thermal current. Therefore, unfavorable values of $q_{cQ}$ can only reduce the enhancement factor down to 1, which is realized only when $q_{cQ} = \frac{q_{cw}}{q_{Qw}}$ or $q_{cQ} = \frac{q_{Qw}}{q_{cw}}$. For the parameters chosen in Fig.2 (b), the enhancement factor has a minimum when $q_{cQ} = \frac{q_{Qw}}{q_{cw}}$ [as illustrated by the triangle point in Fig.2(b)]. Away from this point, the cooperative effect can considerably enhance the optimal energy efficiency. This gives rise to a useful route toward high energy-efficiency: tuning the thermoelectric coupling $q_{cQ}$ to enhance the cooperative effect for high energy-efficiency. Note that this method can be applied to materials with low energy-efficiency for the electrical (or thermal) driving magnetic DW motion, which might be of practical usage.

Finally, we emphasize that although the electrical and thermal currents coexist in most practical situations, the cooperative maximum efficiency can be reached only by properly tuning the temperature and electrochemical potential gradients [37], as shown in Fig.1 (b).

*Cooperative effects in spin-thermoelectric systems.* In a magnetic material the coupled spin, charge and heat transport is described by the following phenomenological equation [26, 32],

$$\begin{pmatrix} \mathcal{J}_c \\ \mathcal{J}_s \\ \mathcal{J}_Q \end{pmatrix} = \begin{pmatrix} G & GP & GST \\ GP & G & GS'T \\ GST & GS'T & K_0 T \end{pmatrix} \begin{pmatrix} \Delta V \\ \Delta m \\ \Delta T/T \end{pmatrix}. \quad (5)$$

where $\mathcal{J}_c = j^{(\uparrow)} + j^{(\downarrow)}$, $\mathcal{J}_s = j^{(\uparrow)} - j^{(\downarrow)}$ with $j^{(\uparrow)}$ and $j^{(\downarrow)}$ denoting the electrical currents of the spin-up and spin-down electrons, respectively. $\Delta V = (\mu_h - \mu_c)/e$ and $\Delta m = m_h - m_c$ with $\mu \equiv (\mu_\uparrow + \mu_\downarrow)/2$, and $m \equiv (\mu_\uparrow - \mu_\downarrow)/(2e)$ where $\mu_\uparrow$ and $\mu_\downarrow$ are the electrochemical potentials for spin-up and spin-down electrons, respectively. $G = \sigma A/l$ is the electrical conductance and $K_0 = \kappa_0 A/l$ is the heat conductance of the device at $\Delta V = \Delta m = 0$ with $\kappa_0$ being the heat conductivity. Possible applications of the system include electrical power generator, cooling/heat-pumping, and spin pumper (the former two are illustrated in Fig. 3).

We first discuss the spin-thermoelectric power generator driven by the coexisting temperature gradient $\Delta T/T$ and spin density gradient $\Delta m$ [Fig. 3(a)]. The energy efficiency is given by $\phi = -\mathcal{I}_c \Delta V/(\mathcal{I}_Q \Delta T/T + \mathcal{I}_s \Delta m)$. Using Eqs. (1) and (5) we obtain

$$\xi = \frac{\kappa_0 P^2 + \sigma T(S^2 - 2PSS')}{\kappa_0(1-P^2) - \sigma T(S^2 - 2PSS' + S'^2)}. \tag{6}$$

Again, the above figure of merit is always *greater* than (or, at least, equal to) both the figure of merit for thermoelectric power generator $\xi_{TE} = \frac{\sigma S^2 T}{\kappa_0 - \sigma S^2 T}$ and the figure of merit for spin-charge conversion $\xi_{SE} = \frac{P^2}{1-P^2}$.

We show in Fig. 4(a) that the enhancement factor of the figure of merit induced by the cooperative effect is considerable when $P$ and $S'(100\mu V/K)$ differs from each other (especially when they have different signs). It is shown in the supplementary material [44] that the transport coefficients are bounded by the second-law of thermodynamics [40]. The white regions in Fig. 4 are forbidden by the second-law of thermodynamics. The cooperative figure of merit $\xi$ can be very large for large $P$ and $S'$ as shown in Fig. 4(b). Particularly, the figure of merit $\xi$ is very large near the boundary of the allowed region. Exactly speaking the boundary represents the limit when the determinant of Onsager matrix becomes zero [i.e., $\kappa_0(1-P^2) - \sigma T(S^2 - 2PSS' + S'^2) = 0$]. Thus, the maximum energy efficiency approaches its upper bound, 100%, and the figure of merit $\xi$ goes to infinity in approaching the boundary of the allowed region in Fig. 4 (The divergent behavior of $\xi$ cannot be resolved in the figure due to restricted data range and resolution). We also plot the enhancement factor as a function of $P$ and the thermoelectric degree of coupling, $\lambda_{TE} = \frac{\sigma S^2 T}{\kappa_0}$, for $S' = 25\mu V/K$ in Fig. 4(c). From Fig. 4(c) the enhancement of energy efficiency due to cooperative effects is considerably large for large $|P|$ and $\lambda_{TE}$. The enhancement factor as a function of the spin-Seebeck coefficient $S'$ and the thermoelectric degree of coupling $\lambda_{TE}$ for $P = 0.5$ is plotted in Fig. 4(d). We found that the enhancement is considerable for negative $S'$ with large $|S'|$. That is, strong enhancement can be obtained when $S' < 0$ for moderate and small thermoelectric degree of coupling $\lambda_{TE}$. The negative $S'$ and positive $S$ require that the Seebeck coefficient of the minority-spin carriers to be negative (with large absolute value) while the majority-spin carriers have positive Seebeck coefficient.

We now consider the spin-thermoelectric cooling/heating driven by the coexisting voltage $\Delta V$ and spin chemical potential gradient $\Delta m$. The coefficient of performance of the refrigerator (and heat pumper) is defined as $\eta \equiv \frac{\dot{Q}}{\dot{W}} = \frac{T}{\Delta T} \frac{-J_Q \Delta T/T}{J_c \Delta V + J_s \Delta m} = \eta_c \phi$, where $\eta_c = \frac{T}{\Delta T}$ is the Carnot efficiency. The schematic of spin-thermoelectric cooling/heating is shown in Fig. 3(b) and here we discuss cooling as an example. Using Eqs. (1) and (5), we obtain

$$\xi = \frac{\sigma T \left(S^2 - 2PSS' + S'^2\right)}{\kappa_0 (1-P^2) - \sigma T \left(S^2 - 2PSS' + S'^2\right)}. \tag{7}$$

The above figure of merit is *greater* or equal to both the figure of merit for thermoelectric cooling, $\xi_{TE}$, and the figure of merit for the spin-Peltier cooling $\xi_{SP} = \frac{\sigma T S'^2}{\kappa_0 - \sigma T S'^2}$ [51].

The enhancement factor of the figure of merit induced by cooperative effect is plotted in a wide parameter range in Fig. 5(a). The figure of merit is significantly improved by cooperative effect when $P$ strongly deviates from $S'/(100 \mu V/K)$ (particularly when the two have opposite signs). From Fig. 5(b), one can see that the cooperative figure of merit can be much larger than 1 when $P$ and $S'/(100\mu V/K)$ are sufficiently different. In such a regime, the spin-thermoelectric refrigeration is much more efficient than the thermoelectric cooling in the same material. Particularly, when $S'/(100\mu V/K)$ is close to -0.5, the spin-thermoelectric refrigeration becomes more efficient than the thermoelectric refrigeration. Figs. 5(c) and 5(d) indicate that the enhancement of figure of merit is strong when $P$ and $S'/(100\mu V/K)$ are very different, as shown by Fig. 5(b) and 5(d).

*Conclusion and discussions.* We have shown that cooperative effects can be a potentially useful tool in improving the energy efficiency of spin caloritronic devices. For example, the cooperative effect can greatly enhance the figure of merit, if the magnetic DW motion is driven by the electrical and heat currents concurrently. Our scheme provides a new route to significantly enhance the energy efficiency and hence considerably reduce Joule heating for future advanced information storage and information processing based on magnetic materials.

*Acknowledgments.* J.H.J acknowledges supports from the faculty start-up funding of Soochow University. He thanks Prof. Gerrit E. W. Bauer for many helpful and stimulating discussions. F.Z and Z.Z.S acknowledge supports from the National Natural Science Foundation of China (NSFC Grant No. 11274236) and Research Fund for the Doctoral Program of Higher Education of China (RFDP Grant No. 20123201110003).

∗ phzzsun@suda.edu.cn

[#] joejhjiang@sina.com

[1] D. A. Allwood, G. Xiong, C. C. Faulkner, D. Atkinson, D. Petit, and R. P. Cowburn, *Science* **309**, 1688 (2005).

[2] R. P. Cowburn, *Nature* (London) **448**, 544 (2007).

[3] D. Atkinson, D. A. Allwood, G. Xiong, M. D. Cooke, C. Faulkner, and R. P. Cowburn, *Nat. Mater.* **2**, 85 (2003).

[4] S. S. P. Parkin, M. Hayashi, and L. Thomas, *Science* **320**, 190 (2008).

[5]. N. Romming, C. Hanneken, M. Menzel, J. E. Bickel, B. Wolter, K. v. Bergmann, A. Kubetzka, and R. Wiesendanger, *Science* **341**, 636 (2013).

[6] J. Slonczewski, *J. Magn. Magn. Mater.* **159**, L1 (1996); L. Berger, *Phys. Rev. B* **54**, 9353 (1996).

[7] Z. Li and S. Zhang, *Phys. Rev. Lett.* **92**, 207203 (2004).

[8] G. Tatara and H. Kohno, *Phys. Rev. Lett.* **92**, 086601 (2004).

[9] G. S. D. Beach, C. Knutson, C. Nistor, M. Tsoi, and J. L. Erskine, *Phys. Rev. Lett.* **97**, 057203 (2006).

[10] S. Emori, U. Bauer, S. M. Ahn, E. Martinez, and G. S. D. Beach, *Nat. Mater.* **12**, 611 (2013).

[11] K.-S. Ryu, L. Thomas, S.-H. Yang, and S. Parkin, *Nat. Nanotechnol.* **8**, 527 (2013).

[12] J. Iwasaki, M. Mochizuki and N. Nagaosa, *Nat. Commun.* **4**, 1463 (2013); J. Iwasaki, M. Mochizuki and N. Nagaosa, *Nat. Nanotechnol.* **8**, 742 (2013); N. Nagaosa and Y. Tokura, *Nat. Nanotechnol.* **8**, 899 (2013).

[13] X. Z. Yu, N. Kanazawa, W. Z. Zhang, T. Nagai, T. Hara, K. Kimoto, Y. Matsui, Y. Onose and Y. Tokura, *Nat. Commun.* **3**, 988 (2012).

[14] L. Thomas, R. Moriya, C. Rettner and S. Parkin, *Science* **330**, 1810 (2010).

[15] M. Hatami, G. E. W. Bauer, Q. Zhang, and P. J. Kelly, *Phys. Rev. Lett.* **99**, 066603 (2007).

[16] A. A. Kovalev and Y. Tserkovnyak, *Phys. Rev. B* **80**, 100408 (2009).

[17] D. Hinzke and U. Nowak, *Phys. Rev. Lett.* **107**, 027205 (2011).

[18] P. Yan, X. S. Wang, and X. R. Wang, *Phys. Rev. Lett.* **107**, 177207 (2011); P. Yan and G. E.W. Bauer, *Phys. Rev. Lett.* **109**, 087202 (2012); P. Yan, Y. Cao, and J. Sinova, *Phys. Rev. B* **92**, 100408 (2015).

[19] A. A. Kovalev and Y. Tserkovnyak, *Europhys. Lett.* **97**, 67002 (2012).

[20] X. S. Wang and X. R. Wang, *Phys. Rev. B* **90**, 014414 (2014).

[21] G. Tatara, *Phys. Rev. B* **92**, 064405 (2015).

[22] W. Jiang, P. Upadhyaya, Y. Fan, J. Zhao, M. Wang, L.-T. Chang, M. Lang, K. L. Wong, M. Lewis, Y.-T. Lin, J. Tang, S. Cherepov, X. Zhou, Y. Tserkovnyak, R. N. Schwartz, and K. L. Wang, *Phys. Rev. Lett.* **110**, 177202 (2013).

[23] L. Kong and J. Zang, *Phys. Rev. Lett.* **111**, 067203 (2013).

[24] M. Mochizuki, X. Z. Yu, S. Seki, N. Kanazawa, W. Kochibae, J. Zang, M. Mostovoy, Y. Tokura, and N. Nagaosa, *Nat. Mater.* **13**, 241 (2014).

[25] A. A. Kovalev, *Phys. Rev. B* **89**, 241101 (2014).

[26] G. E. W. Bauer, E. Saitoh, and B. J. van Wees, *Nat. Mater.* **11**, 391 (2012).

[27] Robert L. Stamps, Stephan Breitkreutz, Johan Akerman, Andrii V. Chumak, YoshiChika Otani, Gerrit E. W. Bauer, Jan-Ulrich Thiele, Martin Bowen, Sara A. Majetich, Mathias Kläui, Ioan Lucian Prejbeanu, Bernard Dieny, Nora M. Dempsey, Burkard Hillebrands, *J. Phys. D: Appl. Phys.* **47**, 333001 (2014).

[28] S. R. Boona, R. C. Myers and J. P. Heremans, *Energy Environ. Sci.* **7**, 885 (2014).

[29] A. Hoffmann and S. D. Bader, *Phys. Rev. Applied* **4**, 047001 (2015).

[30] I. J. Acosta, M. A. O. Robles, S. Bosu, Y. Sakuraba, T. Kubota, S. Takahashi, K. Takanashi and G. E. W. Bauer, arXiv:1506.02777v1.

[31] K. Shen and G. E.W. Bauer, *Phys. Rev. Lett.* **115**, 197201 (2015)

[32] Adam B. Cahaya, Oleg A. Tretiakov, G. E. W. Bauer, *IEEE Trans. Magn.* **51**, 0800414 (2015); Adam B. Cahaya, Oleg A. Tretiakov, G. E. W. Bauer, *Appl. Phys. Lett.* **104**, 042402 (2014).

[33] A. Kirihara, K. Uchida, Y. Kajiwara, M. Ishida, Y. Nakamura, T. Manako, E. Saitoh, and S. Yorozu, *Nat. Mater.* **11**, 686 (2012).

[34] Tianjun Liao, Jian Lin, Guozhen Su, Bihong Lin and Jincan Chen, *Nanoscale*, **7**, 792 (2015).

[35] J. Brüggemann, S. Weiss, P. Nalbach, and M. Thorwart, *Phys. Rev. Lett.* **113**, 076602 (2014).

[36] N. N. Mojumder, D. W. Abraham, K. Roy, and D. C. Worledge, *IEEE Trans. Magn.* **48**, 2016 (2012).

[37] Jian-Hua Jiang, *Phys. Rev. E* **90**, 042126 (2014).

[38] L. Onsager, *Phys. Rev.* **37**, 405 (1931); **38**, 2265 (1931).

[39] S. R. De Groot and P. Mazur, *Non-Equilibrium Thermo-dynamics*, (North-Holland, Amsterdam, 1984).

[40] L. D. Landau and E. M. Lifshitz, *Statistical Physics, part I* (Pergamon, 1958), chap. 12.

[41] G. Nicolis, *Rep. Prog. Phys.* **42**, 225 (1979).

[42] A. De Vos, *J. Phys. Chem.* **95**, 4534 (1991).

[43] J. Yvon, in *Proceedings International Conference on Peaceful Uses of Atomic Energy* (United Nations, Geneva, 1955), *p. 387*; F. L. Curzon and B. Ahlborn, *Am. J. Phys.* **43**, 22 (1975).

[44] See Supplemental Materials at [] for the derivation of the maximum energy efficiency in systems with symmetric Onsager matrices.

[45] S. R. Caplan, *J. Theor. Biol.* **10**, 209 (1966).

[46] H. T. Odum and R. C. Pinkerton, *Am. Sci.* **43**, 331 (1955).

[47] A. Bejan, *Advanced Engineering Thermodynamics* (John Wiley and Sons, NJ, 2006), chap. 3.

[48] O. Kedem and S. R. Caplan, *Trans. Faraday Soc.* **61**, 1897 (1965).

[49] C. Van den Broeck, *Europhys. Lett.* **101**, 10006 (2013); B. Gaveau, M. Moreau, and L. S. Schulman, *Phys. Rev. Lett.* **105**, 060601 (2010); B. Gaveau, M. Moreau, and

L. S. Schulman, *Phys. Rev. E* **82**, 051109 (2010); U. Seifert, *Phys. Rev. Lett.* **106**, 020601 (2011); U. Seifert, *Rep. Prog. Phys.* **75**, 126001 (2012).

[50] Gerrit E. W. Bauer, Stefan Bretzel, Arne Brataas, and Yaroslav Tserkovnyak, *Phys. Rev. B* **81**, 024427 (2010).

[51] J. Flipse, F.K. Dejene, D. Wagenaar, G. E. W. Bauer, J. Ben Youssef, and B. J. van Wees, *Phys. Rev. Lett.* **113**, 027601 (2014).

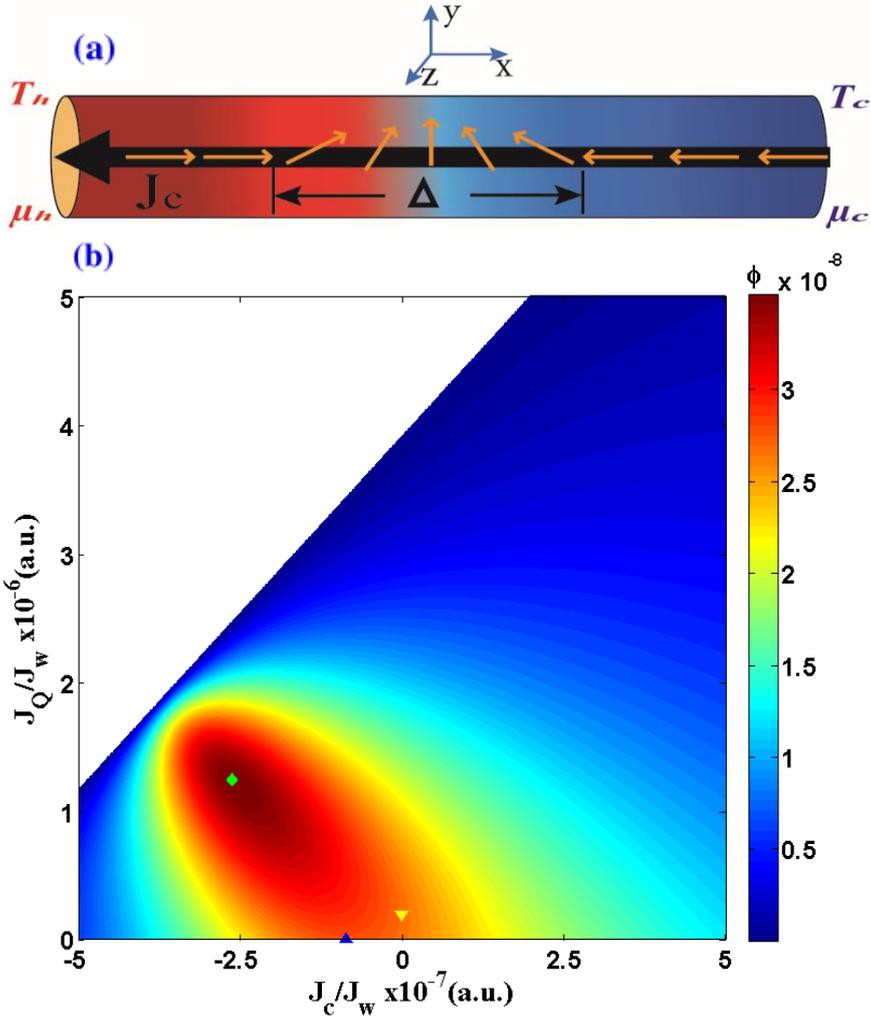

FIG. 1. (Color online) (a) Schematic of electrical and thermal currents induced a 1D head-to-head DW motion in a ferromagnetic nanowire. (b) The energy efficiency, $\phi$, as a function of the ratio of the input currents and output current, $J_c/J_w$ and $J_Q/J_w$. The parameters are $P = 0.5$, $S = 100 \mu V/K$ and $S' = -80 \mu V/K$. The device does not work as current-driven DW motion function in the white region.

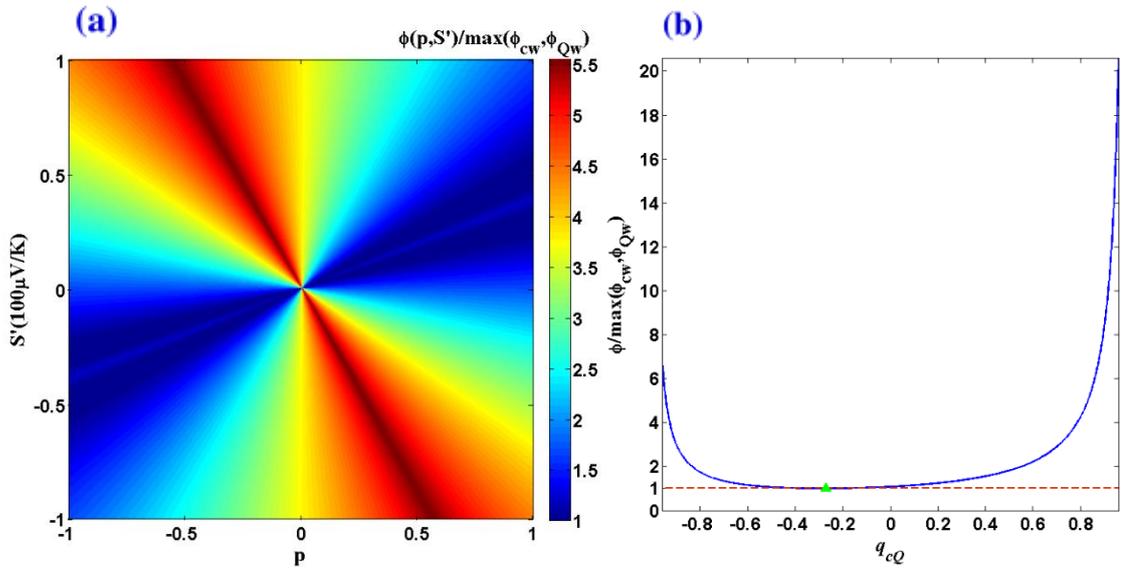

FIG. 2. (Color online) (a) The enhancement factor of the maximum energy efficiency due to the cooperative effect, $\frac{\phi}{max(\phi_{cw},\phi_{Qw})}$, as a function of $P$ and $S'$. (b) The enhancement factor of the maximum energy efficiency as a function of the thermoelectric coupling coefficient $q_{cQ}$. The red dashed line represents the condition with $\frac{\phi}{max(\phi_{cw},\phi_{Qw})} = 1$, which is reached at the green triangle point.

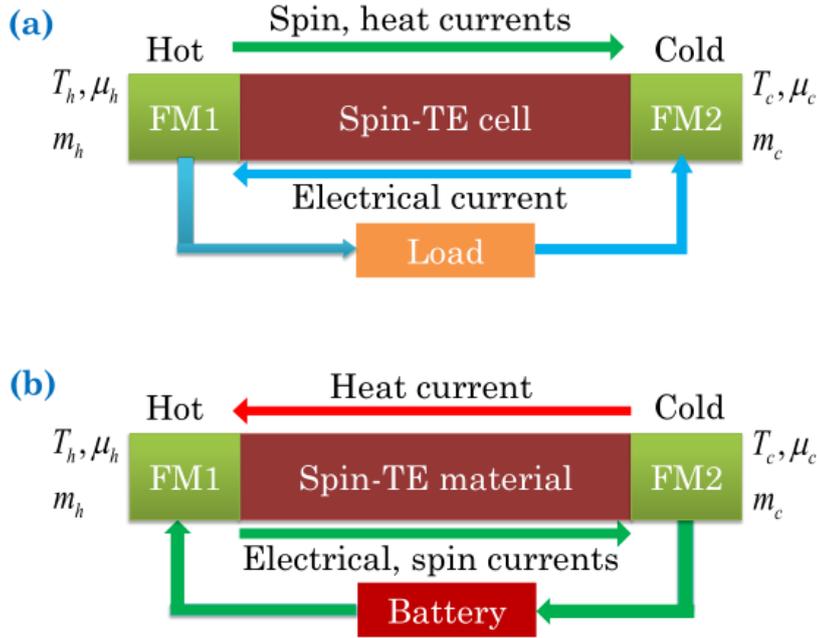

FIG. 3. (Color online) Schematic of (a) the spin-thermoelectric power generator and (b) the spin-thermoelectric cooling/heat-pumper. A spin-thermoelectric ("spin-TE") material (i.e., a conducting magnetic material) sandwiched between two ferromagnetic (FM) electrodes with different temperature, $T_h > T_c$, where the subscripts $h$ and $c$ denoting the hot and cold terminals, respectively.

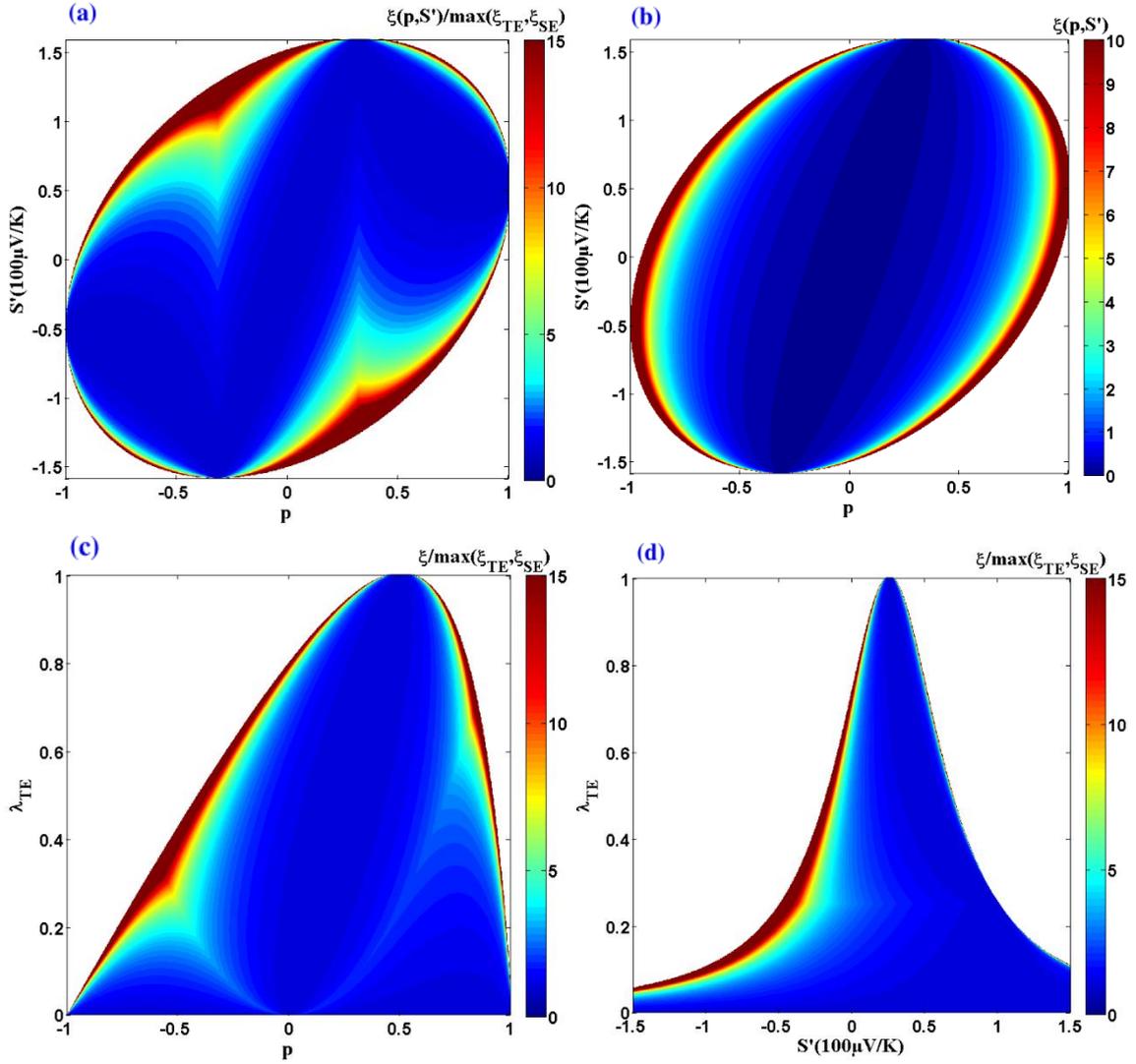

FIG. 4. (Color online) Spin-thermoelectric power generation. (a) The enhancement factor of the figure of merit due to cooperative effects, $\xi/max(\xi_{TE},\xi_{SE})$, as a function of $P$ and $S'$. The parameters are $S = 50\mu V/K$ and $T = 300K$. The heat conductivity is $\kappa_0 = \sigma LT$ with the Lorenz number of $L = 2.5 \times 10^{-8}\ W\Omega K^{-2}$. (b) The cooperative figure of merit $\xi$ as a function of $P$ and $S'$. (c) The enhancement factor of the figure of merit as a function of $P$ and $\lambda_{TE} = \frac{\sigma S^2 T}{\kappa_0}$ for $S' = 25\mu V/K$. (d) The enhancement factor of the figure of merit as a function of $S'$ and $\lambda_{TE} = \frac{\sigma S^2 T}{\kappa_0}$ for $P = 0.5$. In all of the above figures, the white regions are forbidden by the second-law of thermodynamics.

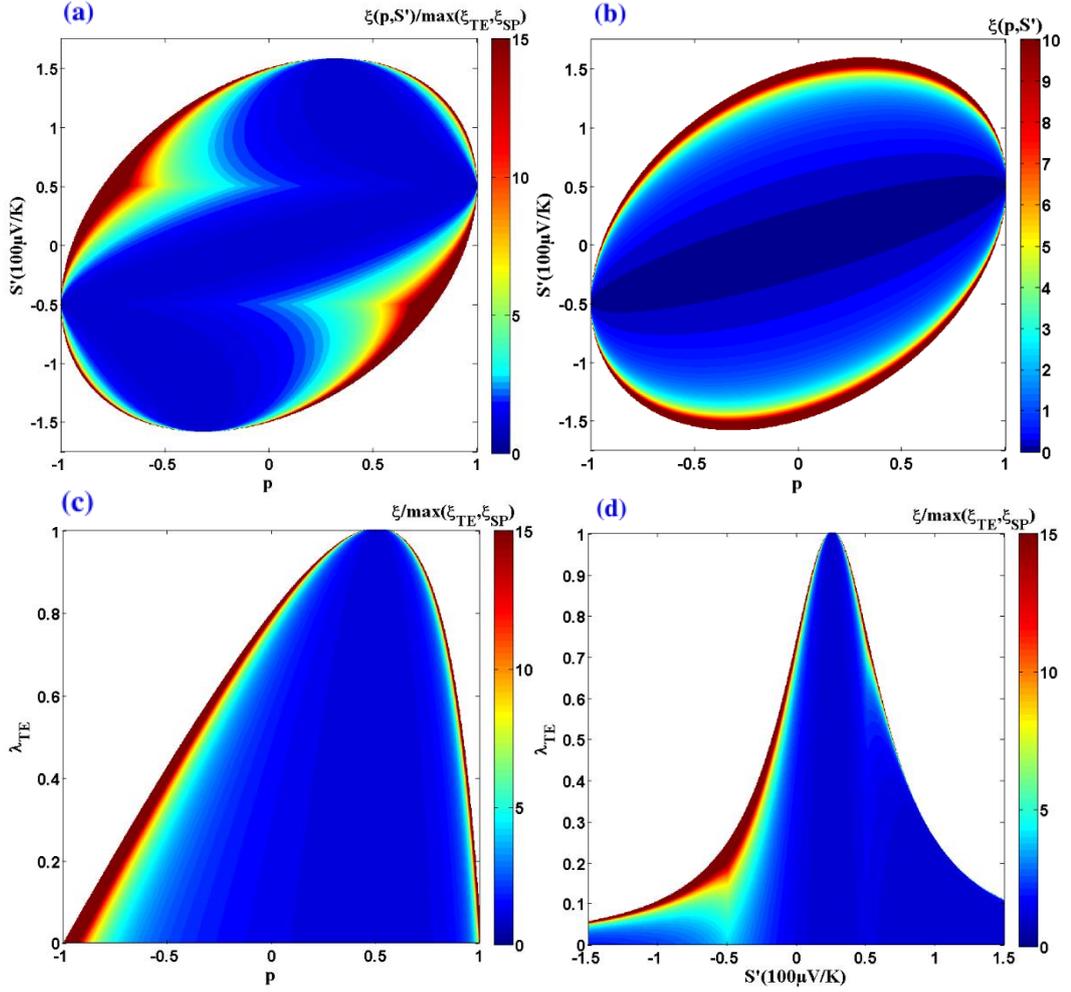

FIG. 5. (Color online) Spin-thermoelectric cooling. (a) The enhancement factor of the figure of merit due to cooperative effects, $/max(\xi_{TE}, \xi_{SP})$ , as a function of $P$ and $S'$. The parameters are $S = 50\mu V/K$ and $T = 300K$. The heat conductivity is $\kappa_0 = \sigma LT$ with the Lorenz number of $L = 2.5 \times 10^{-8} \, W\Omega K^{-2}$. (b) The cooperative figure of merit $\xi$ as a function of $P$ and $S'$. (c) The enhancement factor of fthe igure of merit as a function of $P$ and $\lambda_{TE} = \frac{\sigma S^2 T}{\kappa_0}$ for $S' = 25\mu V/K$. (d) The enhancement factor of the figure of merit as a function of $S'$ and $\lambda_{TE} = \frac{\sigma S^2 T}{\kappa_0}$ for $P = 0.5$. In all of the above figures, the white regions are forbidden by the second-law of thermodynamics.